# The influence of Coulomb interactions on electrical conduction through short molecular wires

Kamil Walczak

Institute of Physics, Adam Mickiewicz University
Umultowska 85, 61-614 Poznań, Poland

Electrical conduction through a two-terminal molecular device is studied using non-equilibrium Green's function (NEGF) formalism. Such junction is made of a short linear wire which is connected to the metallic electrodes. Molecule itself is described with the help of Hückel (tight-binding) model with the electron interactions treated within extended Hubbard model (EHM), while the coupling to the electrodes is described with the help of a broad-band theory. Coulomb interactions within molecular wire are treated by means of the restricted Hartree-Fock (RHF) approximation. In particular, the influence of short-range and long-range Coulomb interactions on electrical transport characteristics is discussed in detail.

## I. Introduction

Molecular electronics attracts much attention of researches and engineers, offering a possibility to overcome some difficulties in further device miniaturization. There is a strong belief that molecular junctions, as composed of molecules connected to macroscopic electrodes, could play the role of active components in future electronic circuits [1-3]. However, such devices are important both from technological as well as from fundamental viewpoint, since they offer a challenging opportunity to understand the effects of many-body interactions on quantum transport through small systems. In reality, several experiments based on scanning tunneling microscopy (STM) [4-6] or on the method of mechanically controllable break-junctions (MCB) [7-9] were performed to obtain molecular junctions and to measure their current-voltage (I-V) characteristics. A full knowledge of the conduction mechanism at the molecular scale is not completed yet, but transport properties of considered structures are associated with some quantum effects, such as: quantum tunneling process, quantization of molecular energy levels, discrete nature of electron charge and spin. In general, the current flowing through the molecular junction is influenced by: the internal structure of molecular system, electronic properties of the electrodes near Fermi energy level, the strength of the molecule-to-electrodes coupling and voltage drop across the molecular wire in the presence of applied bias.

    Theoretical studies presented in this work are focused on the influence of Coulomb interactions (short-rage and long-range) on the electrical conduction through short molecular wires. Different types of nanowires can be joined to metallic electrodes in order to fabricate molecular devices. Since organic conjugated molecules consisting of carbon atoms have delocalised pi-electrons which are suitable for conduction, they are usually suggested as a good candidates for molecular bridges [1]. It should be mentioned that linear carbon-atom chains containing up to $N = 20$ atoms connected at the ends to metal atoms have been

synthesized [10] and recognized as ideal one-dimensional wires [11]. Our computational scheme is based on non-equilibrium Green's functions (NEGF) formalism [12-14], where molecular wire is described with the help of Hückel (tight-binding) model with the electron-electron interactions treated within extended Hubbard model (EHM), while the coupling to the electrodes is described with the help of a broad-band theory. Coulomb interactions within the molecule are treated by means of a restricted Hartree-Fock (RHF) approximation. Anyway, proposed model can be used to simulate organic molecules connected to broad-band metals by properly matching of model parameters. Formalism presented in this work can be also used to study the current in the case of spin-dependent transport, which is of much interest in the study of spintronics and single spin detection.

## II. Computational details

The Hamiltonian of the entire system composed of two electrodes spanned by a molecular wire can be expressed as a three-part sum:

$$H_{tot} = H_{el} + H_{mol} + H_{el-mol}. \tag{1}$$

The first term describes the metallic electrodes which are treated as reservoirs of non-interacting electrons:

$$H_{el} = \sum_{k,\sigma \in \alpha} \varepsilon_{k,\sigma} n_{k,\sigma}, \tag{2}$$

Here: $\varepsilon_{k,\sigma}$ is the single-particle energy of conduction electrons, $n_{k,\sigma} \equiv c_{k,\sigma}^+ c_{k,\sigma}$ denotes the number operator for electrons with momentum $k$ and spin $\sigma$, while $\alpha (= L/R)$ stands for the case of left/right (source/drain) electrode, respectively. In the presence of bias voltage, energies $\varepsilon_{k,\sigma}$ are shifted in the following way:

$$\varepsilon_{k,\sigma} \to \varepsilon_{k,\sigma} + \eta eV, \tag{3a}$$

$$\varepsilon_{k,\sigma} \to \varepsilon_{k,\sigma} - (1-\eta)eV \tag{3b}$$

respectively in the left and right electrodes, while chemical potentials of that electrodes are defined through the following relations:

$$\mu_L = \varepsilon_F + \eta eV, \tag{4a}$$

$$\mu_R = \varepsilon_F - (1-\eta)eV. \tag{4b}$$

Here: $\varepsilon_F$ denotes the equilibrium Fermi level, while $\eta = t_R/(t_R + t_L)$ is the voltage division factor [5,15] which models the potential drop at the contacts ($t_\alpha$ is hopping integral responsible for the strength of the coupling with the $\alpha$ electrode). The second term represents a linear N-atom chain, which is described within the extended Hubbard model [16,17], where not only on-site Coulomb interactions are taken into account but also inter-site ones:

$$H_{mol} = \sum_{i,\sigma} \varepsilon_{i,\sigma} n_{i,\sigma} - \sum_{i,j,\sigma} [(\beta_{i,j} + (-1)^i \delta) c_{i,\sigma}^+ c_{j,\sigma} + h.c.] + \sum_i U_i n_{i\uparrow} n_{i,\downarrow} + \sum_{i,j,\sigma} W_{i,j} n_{i,\sigma} n_{j,\sigma}. \tag{5}$$

Here: $\varepsilon_{i,\sigma}$ is local site energy, $\beta_{i,j}$ is inter-site hopping integral, $U_i$ is on-site Coulomb interaction between two electrons with opposite spins, $W_{i,j}$ is inter-site Coulomb interaction parameters, while $n_{i,\sigma} \equiv c_{i,\sigma}^+ c_{i,\sigma}$, $c_{i,\sigma}^+$ and $c_{i,\sigma}$ denote the number, creation and annihilation operators for electron on site $i$ with spin $\sigma$, respectively. Moreover, it should be noted that summations are restricted to nearest-neighbour sites only, where dimerization (i.e. distinction between double and single bonds) is taken into account by the so-called bond length alternation parameter $\delta$. Setting $W = 0$ in Eq.5 one reproduces Hubbard Hamiltonian, while additionally setting $U = 0$ one obtains Hückel (tight-binding) Hamiltonian. The initial condition stems from the assumption that the electric potential between the electrodes varies linearly with the distance (ramp model) [18]. Therefore, the local site energies $\varepsilon_{i,\sigma}$ are shifted due to this voltage ramp:

$$\varepsilon_{i,\sigma} \to \varepsilon_{i,\sigma} + eV\left[\eta - \frac{i}{N+1}\right]. \tag{6}$$

Furthermore, the third term in Eq.1 corresponds to the tunneling process from the electrodes onto the molecule:

$$H_{el-mol} = \sum_{k,\sigma \in \alpha} t_\alpha \left[c_{k,\sigma}^+ c_{i,\sigma} + h.c.\right]. \tag{7}$$

In our calculations, all the values of energy integrals ($\varepsilon_{i,\sigma} \equiv \varepsilon$, $\beta_{i,j} \equiv \beta$, $U_i \equiv U$, $W_{i,j} \equiv W$ and $t_\alpha$) are treated as parameters which can be reasonably modified.

Since we use Landauer approach to calculate transport characteristics, we solve molecular Hamiltonian (Eq.5) in the restricted Hartree-Fock (RHF) approximation. The averaged (mean-field) form of the Hamiltonian can be written as:

$$\begin{aligned}H_{mol}^{RHF} &= \varepsilon \sum_{i,\sigma} n_{i,\sigma} - \beta \sum_{i,j,\sigma}\left[c_{i,\sigma}^+ c_{j,\sigma} + h.c.\right] - \delta \sum_{i,j,\sigma}\left[(-1)^i c_{i,\sigma}^+ c_{j,\sigma} + h.c.\right] \\ &+ U \sum_{i,\sigma}\langle n_{i,\sigma}\rangle n_{i,\bar{\sigma}} + W \sum_{i,j,\sigma}\langle n_{i,\sigma} + n_{i,\bar{\sigma}}\rangle n_{j,\sigma} - W\sum_{i,j,\sigma}\left[\langle n_{i,\sigma}\rangle c_{i,\sigma}^+ c_{j,\sigma} + h.c.\right]\end{aligned} \tag{8}$$

Here the forth on-site Coulomb repulsion term for spin-up electrons is due to the field created by spin-down electrons, while the fifth inter-site Coulomb repulsion term for spin-up electrons is due to the field created by spin-up as well as spin-down electrons. Both mentioned terms are associated with renormalization of local site energies. The last term in Eq.8 effectively renormalizes hopping integrals of the wire with conservation of spin symmetry. It should be stressed that also unrestricted Hartree-Fock (UHF) approximation with breaking of spin symmetry was used in order to describe organic conjugated molecules [19,20]. Occupation number of the electrons on each site for particular voltages (nonequilibrium case) is determined self-consistently using the Keldysh formalism [12-14]:

$$\langle n_{i,\sigma}\rangle = -\frac{i}{2\pi}\int_{-\infty}^{+\infty} d\omega G_{i\sigma,i\sigma}^<(\omega), \tag{9}$$

Since charge distribution along molecular bridge has influence on molecular parameters ($\varepsilon$ and $\beta$) and such parameters are used later to obtain occupation numbers, we have to recalculate such quantities in the self-consistent procedure until convergence. The whole procedure is repeated for different values of applied bias. It should be also mentioned that potential obtained in this way is only Poissonian contribution to the whole potential, while Laplace part of the potential is included as a ramp model [21]. The lesser Green function $G^<$ can be obtained from the Dyson equation and expressed in the general form as:

$$G^<_{i\sigma,j\sigma} = \sum_{i',j'} G^r_{i\sigma,i'\sigma} \Sigma^<_{i'\sigma,j'\sigma} G^a_{j'\sigma,j\sigma}. \tag{10}$$

The superscripts $r$ and $a$ denote the retarded and advanced Green functions, respectively:

$$G^r(\omega) = \left[1\omega - H^{RHF}_{mol} - \Sigma^r\right]^{-1} \tag{11}$$

and $G^a = [G^r]^*$ (here 1 denotes the unit matrix of the dimension equal to molecular Hamiltonian $N \times N$). The lesser self-energy can be written as follows:

$$\Sigma^<_{i\sigma,j\sigma}(\omega) = i\delta_{i,j}\left[\delta_{i,1}\Gamma_{L\sigma}(\omega)f_L(\omega) + \delta_{i,N}\Gamma_{R\sigma}(\omega)f_R(\omega)\right], \tag{12}$$

where $f_\alpha = f(\omega - \mu_\alpha)$ is Fermi distribution function in the $\alpha$ electrode. The retarded and advanced self-energy functions are given by:

$$\Sigma^r_{i\sigma,j\sigma}(\omega) = \delta_{i,j}\left[\delta_{i,1}\left[\Lambda_{L\sigma}(\omega) - \frac{i}{2}\Gamma_{L\sigma}(\omega)\right] + \delta_{i,N}\left[\Lambda_{R\sigma}(\omega) - \frac{i}{2}\Gamma_{R\sigma}(\omega)\right]\right] \tag{13}$$

and $\Sigma^a = [\Sigma^r]^*$. The real and imaginary terms of the self-energy components are closely connected to each other through the Hilbert transform [18]:

$$\Lambda_{\alpha\sigma}(\omega) = \frac{1}{2\pi} P \int_{-\infty}^{+\infty} dz \frac{\Gamma_{\alpha\sigma}(z)}{\omega - z}, \tag{14}$$

where $P$ is the Cauchy principal value. For the sake of simplicity, in further analysis we make an assumption that local density of states in both electrodes is constant over an energy bandwidth and zero otherwise. In this case, the so-called linewidth function is approximately expressed as:

$$\Gamma_{\alpha\sigma}(\omega) \cong \frac{2t_\alpha^2}{\gamma_{\alpha\sigma}}, \tag{15}$$

where $4\gamma_{\alpha\sigma}$ is the spin-$\sigma$ bandwidth of the $\alpha$ electrode. Such dispersionless coupling to the electrodes is commonly used in the literature and is usually sufficient in describing broad-band metals.

The steady-state electric current flowing through the device can be computed from the time evolution of the occupation number for electrons in the $\alpha$ electrode. Formula for the current can be expressed with the help of the lesser Green function [12-14]:

$$I(V) = -e\frac{d}{dt}\langle N_\alpha\rangle = \frac{e}{h}\sum_{i,k,\sigma} t_\alpha \int_{-\infty}^{+\infty} d\omega\left[G^<_{i\sigma,k\sigma}(\omega) + c.c.\right], \qquad N_\alpha = \sum_{k,\sigma\in\alpha} n_{k,\sigma}. \tag{16}$$

After applying the Dyson equation, Eq.16 can be written through the use of the retarded Green function:

$$I(V) = \frac{e}{h}\sum_\sigma \int_{-\infty}^{+\infty} d\omega [f_L(\omega) - f_R(\omega)]\Gamma_{L\sigma}(\omega)\Gamma_{R\sigma}(\omega)\left|G^r_{1\sigma,N\sigma}(\omega)\right|^2. \tag{17}$$

Finally, the differential conductance is given by the derivative of the current with respect to voltage: $G(V) = dI(V)/dV$. In our calculations, we have assumed that transport process is purely coherent and elastic. It means that the current conservation rule is fulfilled on each site and for any energy $\omega$.

## III. Results, discussion, and final conclusions

Let us consider the situation in which a linear carbon-atom chain containing $N = 4$ atoms (butadiene) is sandwiched between two paramagnetic electrodes. Molecular description itself includes only interacting pi-electrons of hydrocarbons, while the coupling to the electrodes is treated within a broad-band theory. This is a test case simple enough to analyze all the essential physics in detail and to compare obtained results with results already known in the literature [18,22]. Therefore, we take the following energy parameters (given in eV): $\varepsilon = 0$ (the reference energy), $\beta = 2.4$, $\delta = 0.2$, $\Gamma_L = \Gamma_R = 0.4$ (molecule is symmetrically connected to the electrodes). Since the typical metal bandwidth is equal to 10 eV, the strength of the coupling with the electrodes can be estimated as $t_L = t_R \cong 0.7$ eV. The magnitude of Coulomb integrals is dictated by some theoretical studies [23]: $U = 4$ and $W = 2$. Fermi energy for organic junctions is usually fixed in between the highest occupied molecular orbital (HOMO) and the lowest unoccupied molecular orbital (LUMO). In this work we assume that Fermi level is fixed 1 eV below the LUMO level of isolated molecule ($\varepsilon_F = 0.483$). The temperature of the whole system is set at $T = 300$ K.

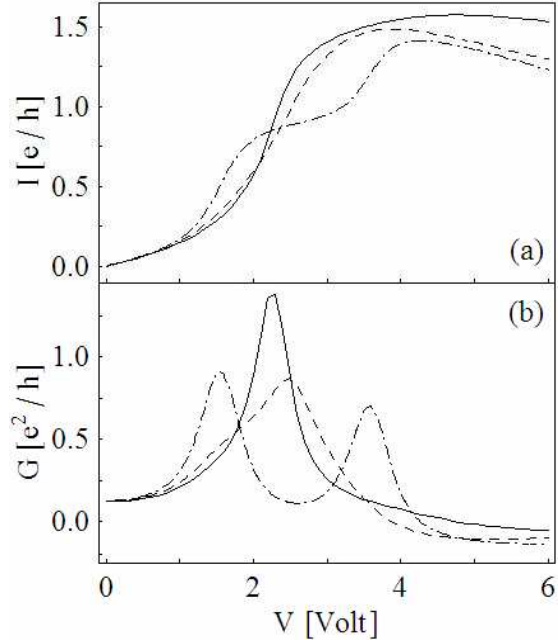

Figure 1: The evolution of (a) $I(V) = -I(-V)$ and (b) $G(V) = G(-V)$ functions for three sets of Coulomb integrals: $U = W = 0$ (dashed-dotted line), $U = 4$ and $W = 0$ (dashed line), $U = 2W = 4$ (solid line).

The numerical results of the $I - V$ dependences for the analyzed four-atom junction are shown in Fig.1a. Computations for $U = W = 0$ (Hückel approach with the potential ramp) represent two distinct current jumps (at $V \cong 1.6$ and $V \cong 3.6$) separated by the current plateau, where the first step is due to the electron conduction through the LUMO level, while the second one is associated with the hole conduction through the HOMO level. Both steps have approximately the same intensity, because of the electron-hole symmetry of the model. Obviously, the current jumps correspond to the conductance peaks, as can be seen in Fig.1b. Moreover, for the case of higher voltages ($V > 4.3$) the calculated transport characteristic exhibits a negative differential conductance (NDC effect). This phenomenon is associated with voltage drop inside the molecule, since it is not observed for the case of flat potential.

Ramp-like potential generates NDC and reduces the magnitude of the current in a high-bias limit (in comparison with results obtained within flat potential model). Both mentioned effects can be justified as a consequence of voltage-induced redistribution of charge carriers along the molecular wire, where in one end of the wire there is depletion while on the other end there is accumulation of electrons (as will be discussed later). In general, the higher applied bias, the stronger polarization effects and therefore the transmission of the current is blocked as voltage is increased for sufficiently large electric fields.

Interactions at the Hubbard level ($U = 4$, $W = 0$) exchanging two steps in the $I-V$ characteristic on one fairly smooth step approximately doubled in height due to the so-called pinning effect, i.e. pinning of the molecular self-consistent levels to electrochemical potentials of the source and drain electrodes [24-26], as illustrated in Fig.1a. Such continuous current curve corresponds to one broadened conductance peak at $V \cong 2.5$, as can be seen in Fig.1b. It means that electron and hole conduction partly merges, where in transport process participates simultaneously both the HOMO and the LUMO levels. Besides, for higher voltages ($V > 4.3$) we can still observe NDC which is present even for unrealistically extreme values of the $U$-parameter and therefore we conclude that short-range interactions (such as on-site Coulomb repulsion) can not reduce this effect. Inclusion of long-range Coulomb interactions ($U = 2W = 4$) sharpens the step in the $I-V$ function (see Fig.1a) or equivalently the peak in the $G-V$ dependence (see Fig.1b) and does not destroy the pinning effect (see Fig.2a). Now, the current jump takes place at $V \cong 2.2$. Such transition in the conduction character (from continuous current curve for Hubbard approach to the jump-like $I-V$ characteristic for EHM) is mainly due to renormalization of site energies. Moreover, in the case of higher voltages ($V > 4.3$) we can observe reduction of NDC, what is a principal consequence of the renormalization of hopping integrals.

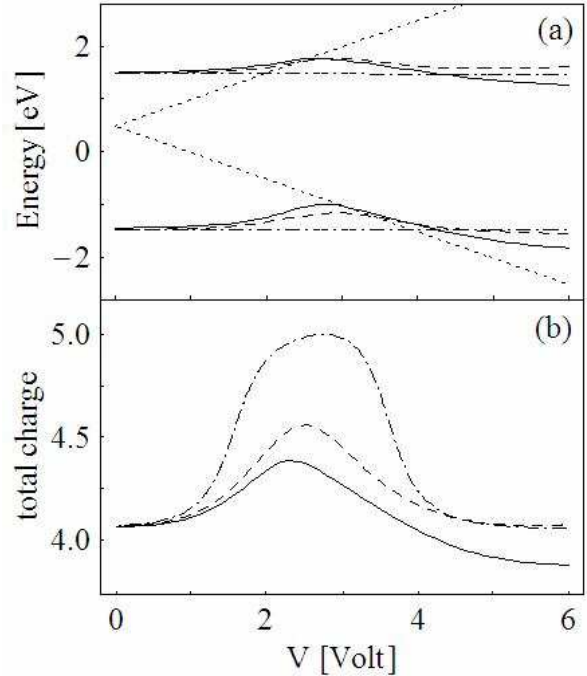

Figure 2: The evolution of LUMO and HOMO levels (a) and total charge occupation (b) for a four-atom junction as a function of applied bias. Figures are plotted for three sets of Coulomb integrals: $U = W = 0$ (dashed-dotted line), $U = 4$ and $W = 0$ (dashed line), $U = 2W = 4$ (solid line), respectively. Electrochemical potentials are denoted by dotted lines.

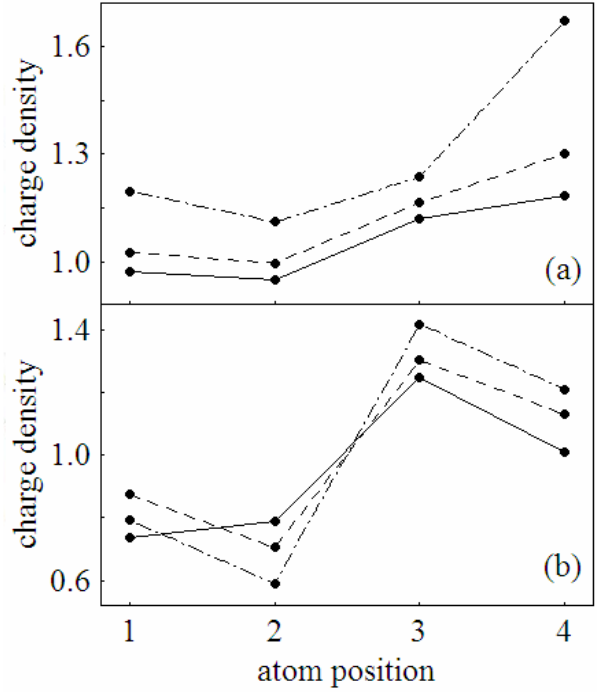

Figure 3: Charge occupation of particular sites for the following bias voltages: $V = 3$ (a) and $V = 6$ (b) for three sets of Coulomb integrals: $U = W = 0$ (dashed-dotted line), $U = 4$ and $W = 0$ (dashed line), $U = 2W = 4$ (solid line), respectively.

    The electron-electron interactions are mainly responsible for neutralizing polarization effect caused by ramp electrostatic potential. The final result is due to the combined effect of voltage-induced redistribution of charge carriers inside the molecule and Coulomb repulsion between electrons. The electrostatic potential spatial profile in molecular devices is a quantity of great interest, since it carries nontrivial information concerning: the molecular screening, and the presence of impurities as well as Schottky-like barriers [20,24-28]. As is evident from Fig.2b, on-site Coulomb interactions strongly reduce an amount of charge carriers on the wire in the voltage-window $1.6 < V < 3.6$ in comparison with noninteracting system, while inter-site interactions produce further reduction of total charge even for $V > 4.3$. Such tendency results from repulsive character of interactions between electrons which are trying to avoid each others within the molecule. Self-consistent charging under bias may also lead to a spatially asymmetric potential profile (or charge distribution along the wire, as shown in Fig.3), although the $I - V$ dependence remains symmetric with respect to bias polarity. This is due to the fact that charging-induced potential asymmetry reverses perfectly on reversing bias, so it can not produce any rectification effect. However, it should be mentioned that the origin of small rectification effect can be associated with the breakdown of the electron-hole symmetry in the case of unrestricted Hartree-Fock (UHF) approach, as discussed in [19].

    Furthermore, the electrostatic potential profile exhibits also the so-called Friedel oscillations that are partially reflected by charge distribution along the wire, where we can observe charge accumulation and charge depletion regions. Since the Friedel oscillation wavelength is about twice as large as inter-atomic spacing for the case of organic molecules [25], in four-atom device we expect about two oscillation circles – what is indeed observed for a suitable high voltages (see Fig.3b). The origin of such oscillations is associated with strong defects of wire-metal contacts. On-site Hubbard interactions reduce Friedel

oscillations, while inclusion of long-range Coulomb interactions leads to further reduction and breaks point symmetry of charge distribution. It should be noted that the amplitudes of such oscillations are also reduced in the presence of inelastic or incoherent processes [25].

Concluding, we have shown that transport at molecular scale is strongly determined by Coulomb interactions between electrons. In particular, the following charging-induced phenomena were observed: mergence of electron and hole conductions due to the discussed pinning effect, NDC generated by long-range Coulomb interactions due to the redistribution of the charges on the molecule, and Friedel oscillations reflected by charge distribution along the molecular wire due to the strong defects of wire-metal contacts.